%% file: main.tex
\documentclass[twocolumn,aps,prd,superscriptaddress,nofootinbib]{revtex4-2}
\usepackage{graphicx}
\usepackage{color}
\graphicspath{{figures/}{fig/}}
\usepackage{amsmath}
\usepackage{amssymb}
\usepackage{bm}
\usepackage{slashed}
\usepackage{epsfig}
\usepackage{amsfonts}
\usepackage{epstopdf}
\usepackage{hyperref}
\usepackage{bbm}
\usepackage{textcomp}
\usepackage{color}

\newcommand{\sect}[1]{\section{#1}}
\newcommand{\mi}{\mathrm{i}}
\newcommand{\md}{\mathrm{d}}

\begin{document}
\title{Efficient Computation of One-Loop Feynman Integrals and Fixed-Branch Integrals to High Orders in $\epsilon$
}

\author{Rui-Jun Huang}
\email{huangrj2215@pku.edu.cn}
\affiliation{School of Physics, Peking University, Beijing 100871, China}

\author{Dong-Shan Jian}
\email{dsjian@stu.pku.edu.cn}
\affiliation{School of Physics, Peking University, Beijing 100871, China}

\author{Yan-Qing Ma}
\email{yqma@pku.edu.cn}
\affiliation{School of Physics, Peking University, Beijing 100871, China}
\affiliation{Center for High Energy Physics, Peking University, Beijing 100871, China}

\author{Dao-Ming Mu}
\email{mudaoming2879@pku.edu.cn}
\affiliation{School of Physics, Peking University, Beijing 100871, China}

\author{Wen-Hao Wu}
\email{whwphys97@outlook.com}
\affiliation{School of Physics, Peking University, Beijing 100871, China}

\date{\today}

\begin{abstract}
We propose a novel method, called the dimension-changing transformation (DCT), to compute one-loop Feynman integrals and recently introduced fixed-branch integrals to arbitrary orders in $\epsilon$. The DCT relates one-loop Feynman integrals or fixed-branch integrals in one spacetime dimension to their corresponding quantities with auxiliary mass in any other dimension, making the expansion to high orders in $\epsilon$ highly efficient. We applied this method to several examples to demonstrate its validity and efficiency. The approach introduced in this work has been implemented in an open-source C++ package, available at \href{https://gitlab.com/multiloop-pku/dct}{https://gitlab.com/multiloop-pku/dct}.
\end{abstract}

\maketitle
\allowdisplaybreaks

\sect{Introduction}
As experimental uncertainties continue to decrease at high-energy colliders such as the LHC, it is essential to control theoretical uncertainties in predictions to a corresponding level. In this context, the computation of Feynman integrals (FIs) plays a crucial role, both for testing the Standard Model of particle physics and for probing new physics. The current standard procedure for evaluating FIs involves first reducing them to a smaller set of basis integrals, known as master integrals (MIs), using techniques such as integration-by-parts identities \cite{Tkachov:1981wb, Chetyrkin:1981qh, Laporta:2000dsw, Gluza:2010ws, Schabinger:2011dz, vonManteuffel:2012np, Lee:2013mka, vonManteuffel:2014ixa, Larsen:2015ped, Peraro:2016wsq, Mastrolia:2018uzb, Liu:2018dmc, Guan:2019bcx, Klappert:2019emp, Peraro:2019svx, Frellesvig:2019kgj, Wang:2019mnn, Smirnov:2019qkx, Klappert:2020nbg, Boehm:2020ijp, Heller:2021qkz, Bendle:2021ueg}, and then calculating these MIs. The computation of one-loop MIs up to the finite part in the dimensional regularization parameter, $\epsilon = (4 - D)/2$, is a well-established problem \cite{tHooft:1978jhc, Passarino:1978jh, Bern:1993kr}, with numerous publicly available packages designed for these calculations \cite{vanOldenborgh:1989wn, Hahn:1998yk, Ellis:2007qk, vanHameren:2010cp, Denner:2016kdg, Cullen:2011kv, Patel:2015tea}. However, for high-precision theoretical predictions, it is necessary not only to compute multi-loop FIs, but also to evaluate higher-order terms in $\epsilon$ for one-loop FIs.

The evaluation of one-loop MIs to higher powers of $\epsilon$ has been studied in several works, such as \cite{Badger:2022mrb,Henn:2022ydo,Buccioni:2023okz}. The auxiliary mass flow (AMFlow) method \cite{Liu:2017jxz,Liu:2021wks,Liu:2022mfb,Liu:2022chg} offers a systematic approach for evaluating both one-loop and multi-loop MIs to arbitrary powers of $\epsilon$ with high precision. However, obtaining higher-order terms in $\epsilon$ often requires solving a series of differential equations at multiple $\epsilon$ values, followed by interpolation, or solving differential equations with $\epsilon$ expanded to a fixed power. In this paper, we demonstrate that, for one-loop MIs, solving differential equations at multiple $\epsilon$ values can be avoided, significantly improving efficiency.

We propose a novel method, termed dimension-changing transformation (DCT), for efficiently computing one-loop Feynman integrals and fixed-branch integrals (FBIs)\footnote{It was demonstrated in Ref. \cite{FBI} that multi-loop FIs can be expressed as integrals over a small number of branch parameters, with the integrand being a linear combination of FBIs. FBIs share a similar structure with one-loop Feynman integrals, and in fact, one-loop Feynman integrals can be considered as special cases of FBIs with the number of branches equals to one. Once FBIs are computed, the evaluation of multi-loop FIs becomes much easier.}. In this approach, differential equations only needs to be solved at an arbitrary value of $\epsilon$, after which other values of $\epsilon$ or higher powers of $\epsilon$ can be efficiently obtained via the DCT.

The method has been implemented in an open-source C++ package that computes one-loop FIs and FBIs to any desired order in $\epsilon$. The performance of the package is demonstrated through several examples.

\sect{Dimension-Changing transformation}
\subsection{One-loop Feynman integrals}
A dimensionally regularized one-loop FI with an auxiliary mass \cite{Liu:2017jxz} can be expressed as:
\begin{equation}\label{eqs: definition of I with eta}
I^{\Delta}_{\vec{\nu}}(\eta)\equiv\int\frac{\md^D l}{\mi \pi^{D/2}}\prod^N_{\alpha=1}\frac{1}{(\mathcal{D}_\alpha- \eta)^{\nu_\alpha}},
\end{equation}
where $\mathcal{D}_\alpha\equiv (l+q_{\alpha})^2-m_\alpha^2$, $q_\alpha$ are external momenta, and $\Delta=\frac{L D}{2}$. Here $L$ is the number of loops, with $L=1$ for one-loop integrals. The actual FI, without the $\eta$ dependent term, is recovered when $\eta$ approaches zero from the negative imaginary axis in the $\eta$-plane, as dictated by Feynman prescription:
\begin{equation}\label{eqs: recover actual I}
I^\Delta_{\vec{\nu}}\equiv \lim_{\eta\to \mi 0^-} I^\Delta_{\vec{\nu}}(\eta).
\end{equation}
Interestingly, the auxiliary mass terms can also be interpreted as the effects of additional space-time dimensions for the loop momentum. This can be illustrated by rewriting a one-loop integral in $D$ dimensions as:
\begin{equation}\label{eqs: definition of I}
I^\Delta_{\vec{\nu}}=\int\frac{\md^{D-D_0} l_\perp}{ \pi^{(D-D_0)/2}}\int\frac{\md^{D_0} l_\parallel}{\mi \pi^{D_0/2}}\prod^N_{\alpha=1}\frac{1}{(\mathcal{D}^\parallel_\alpha-l_\perp^2)^{\nu_\alpha}}, 
\end{equation}
 where the loop momentum $l$ is split into two components: $l_\parallel$, which lives in $D_0$ dimensions, and $l_\perp$, which is orthogonal to the external momenta. Changing variable $l_\perp^2\to \eta$, integrating over the angular components in the orthogonal space, and defining $\delta=L (D-D_0)/2$, we obtain
\begin{equation}\label{eqs: dimension change1}
I^\Delta_{\vec{\nu}}=\frac{1}{\Gamma(\delta)}\int^\infty_0\md\eta\, \eta^{\delta-1}I^{\Delta-\delta}_{\vec{\nu}}(\eta),
\end{equation}
which is called a DCT. This formula relates one-loop FIs with auxiliary mass in $D_0$ dimensions to one-loop FIs without auxiliary mass in $D$ dimensions. Due to analytic continuation within the dimensional regularization scheme, this expression holds for any values of $D_0$ and $D$.

\begin{figure}[htb]
	\begin{center}        \includegraphics[width=0.8\linewidth]{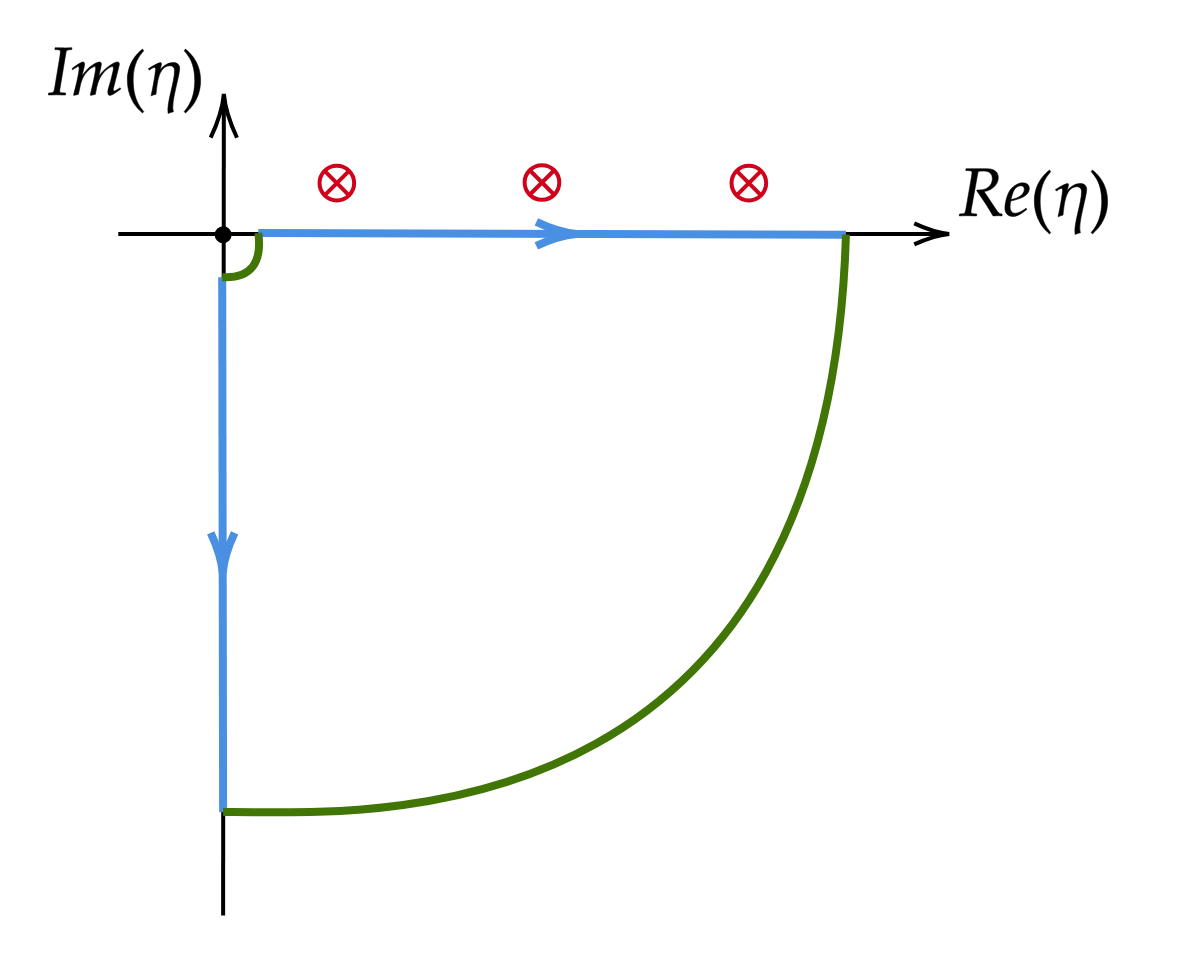}
	\end{center}	\caption{\label{fig:contour}Different choices of integration paths for recovering original FIs.  The green path contributes nothing due to dimensional regularization. The red crosses indicate singularities, while the two blue paths are equivalent, as there are no singularities in the fourth quadrant.}
\end{figure}

A critical challenge in practice is that the DCT in Eq.~\eqref{eqs: dimension change1} is defined along the positive real axis of $\eta$, where singularities may arise for real kinematic configurations. Fortunately, this issue can be circumvented by deforming the integration path, as shown in Fig.~\ref{fig:contour}. This leads to the DCT on the negative imaginary axis:
 \begin{equation}\label{eqs: dimension change2}
I^\Delta_{\vec{\nu}}=\frac{1}{\Gamma(\delta)}\int^{-\mi\infty}_{-\mi 0^+}\md\eta\, \eta^{\delta-1}I^{\Delta-\delta}_{\vec{\nu}}(\eta).
\end{equation}
The above contour choice is valid not only for real kinematic configurations but also for complex configurations, provided that $I^{D_0}_{\vec{\nu}}(\eta)$ does not have poles in the fourth quadrant. Notably, this contour supports Breit-Wigner distribution, where the mass of a propagator is given by $m^2 = M^2 -\mi M \Gamma  $, with $M$ and $\Gamma$ being real and positive.

\subsection{Fixed-branch integrals}
The concept of fixed-branch integrals (FBIs) was introduced in \cite{FBI} by employing Feynman parameters to combine the propagator denominators within the same branch. A general FBI with an auxiliary parameter $\eta$ is defined as:
\begin{equation}
I_{\Vec{\nu}}^{\Delta}(\mathbf{X},\eta)=\frac{(-1)^\nu\Gamma(\nu-\Delta)}{\prod_{\alpha=1}^{N}\Gamma\left(\nu_{\alpha}\right)}\int\left[\mathrm{d}\mathbf{y}\right]\frac{\prod_{\alpha=1}^{N}y_{\alpha}^{\nu_{\alpha}-1}}{\left(\mathcal{F}+\eta\right)^{\nu-\Delta}},
\label{eq:FBI}
\end{equation}
where $\mathbf{X}=(X_1, X_2, \cdots, X_{B-1})$ are branch parameters, $B$ is the number of branches, $\mathcal{F}=\frac{1}{2} \mathbf{y}^T\hspace{-1mm}\cdot\hspace{-1mm} R(\mathbf{X}) \hspace{-0.5mm} \cdot \hspace{-0.5mm} \mathbf{y} $ is the second Symanzik polynomial with $R$ being a symmetric matrix, and the integration measure is:
\begin{equation}
\left[\mathrm{d}\mathbf{y}\right]\equiv \prod_{\alpha=1}^{N}\mathrm{d}y_{\alpha}\prod_{b=1}^{B}\delta\left(1-\sum_{i=1}^{n_b} y_{(b,i)}\right).
\end{equation}
When $B=1$, the FBI $I_{\Vec{\nu}}^{\Delta}(\mathbf{X},\eta)$ becomes  a one-loop FI with auxiliary mass in the Feynman parameter representation.

The desired FBI without auxiliary mass is defined as 
\begin{align}
    I_{\Vec{\nu}}^{\Delta}(\mathbf{X})\equiv \mathcal{I}_{\Vec{\nu}}^{\Delta}(\mathbf{X},\eta\to\mathrm{i}0^-).
\end{align} 
We can derive the DCT formula for FBIs:
\begin{equation}
    \begin{aligned}
        &\int_0^{\infty}\mathrm{d}\eta \;\eta^{\delta-1}I_{\vec{\nu}}^{\Delta}(\mathbf{X},\eta)\\
        &=\frac{(-1)^\nu \Gamma(\nu-\Delta)}{\prod_{\alpha=1}^{N}\Gamma\left(\nu_{\alpha}\right)}\int\left[\mathrm{d}\mathbf{y}\right]\prod_{\alpha=1}^{N}y_{\alpha}^{\nu_{\alpha}-1}\int_0^\infty\mathrm{d}\eta \eta^{\delta-1}\left(\mathcal{F}+\eta\right)^{\Delta-\nu} \\
        &=\frac{(-1)^\nu\Gamma(\nu-\Delta-\delta)\Gamma(\delta)}{\prod_{\alpha=1}^{N}\Gamma\left(\nu_{\alpha}\right)}\int\left[\mathrm{d}\mathbf{y}\right]\prod_{\alpha=1}^{N}y_{\alpha}^{\nu_{\alpha}-1}\mathcal{F}^{\Delta+\delta-\nu}\\
        &=\Gamma(\delta)I_{\vec{\nu}}^{\Delta+\delta}(\mathbf{X}),
    \end{aligned}
\end{equation}
which gives the following result:
\begin{equation}\label{eq:DCT}
\begin{aligned}
     I_{\vec{\nu}}^{\Delta}(\mathbf{X})&=\frac{1}{\Gamma(\delta)}\int_0^{\infty}\mathrm{d}\eta \;\eta^{\delta-1}I_{\vec{\nu}}^{\Delta-\delta}(\mathbf{X},\eta)\\
     &=\frac{1}{\Gamma(\delta)}\int_{-\mathrm{i}0^+}^{-\mathrm{i}\infty}\mathrm{d}\eta \;\eta^{\delta-1}I_{\vec{\nu}}^{\Delta-\delta}(\mathbf{X},\eta),
\end{aligned}
\end{equation}
which is a generalized version of Eq.~\eqref{eqs: dimension change2}.
Thus, to compute an FBI (or its special case, a one-loop FI), we only need to compute its corresponding quantity with auxiliary mass in any nonsingular space-time dimension.

\sect{Computing integrals with auxiliary mass}
The differential equations of FBIs with respect to the auxiliary parameter $\eta$ are given by \cite{FBI}
\begin{equation}
    \begin{aligned}
        &(2z_0 \eta-C)\frac{\mathrm{d}}{\mathrm{d}\eta}I_{\vec{\nu}}^{\Delta}(\mathbf{X},\eta)\\
        = &\left(2\Delta-\nu-B\right)z_0 I_{\vec{\nu}}^{\Delta}(\mathbf{X},\eta)+\sum_{\alpha=1}^{N} z_\alpha I_{\vec{\nu}-\vec{e}_\alpha}^{\Delta-1}(\mathbf{X},\eta),
        \end{aligned}\label{eq:etaDE}
\end{equation}
where $C=\sum_{b=1}^B C_b$. The unknowns are constrained by the condition:
\begin{equation}\label{eq:Cz}
    S \cdot(C_1,\cdots,C_B,z_1,\cdots,z_N)^T=(z_0,\cdots,z_0,0,\cdots,0)^T\,,
\end{equation}
where $S$ is a generalized Gram matrix extended from $R$ according to the following rules:
\begin{itemize}
    \item \(S_{\alpha, \beta} = R_{\alpha-B, \beta-B}\) if \(\alpha > B\) and \(\beta > B\),
\item \(S_{\alpha, \beta} = 1\) if \(\alpha \leq B\) and \(1 \leq \beta - B - \sum_{b = 1}^{\alpha - 1} n_b \leq n_\alpha\), or \(\beta \leq B\) and \(1 \leq \alpha - B - \sum_{b = 1}^{\beta - 1} n_b \leq n_\beta\),
\item \(S_{\alpha, \beta} = 0\) otherwise.
\end{itemize}
If $\det(S)\neq0$, we set $z_0=1$ and then the constants $C_b$ and $z_\alpha$ are fully determined. If $\det(S)=0$, we set $z_0=0$ and then $C_b$ and $z_\alpha$ belong to any basis of the null space of $S$. We then select a basis such that $C$ is as nonzero as possible. Note that, when $B=1$, the differential equation in Eq.~\eqref{eq:etaDE} reduces to the differential equation for one-loop FIs with respect to auxiliary mass, as derived in Ref.~\cite{Liu:2017jxz}.

We find that if $\det(S)$ and $C$ of $I_{\vec{\nu}}^{\Delta}(\mathbf{X},\eta)$ do not vanish simultaneously, while $\det(S)$ and $C$ of $I_{\vec{\nu}-\vec{e}_\alpha}^{\Delta-1}(\mathbf{X},\eta)$ are both zero, then the corresponding $z_{\alpha}$ is zero. Since MIs of FBIs can always be chosen as corner integrals, where the components of $\vec{\nu}$ are either 0 or 1, we can construct a closed system of differential equations with corner FBIs, ensuring that $z_0$ and $C$ do not vanish simultaneously.

The boundary conditions for differential equations can be taken as $\eta\to\infty$. In this limit, based on the definition Eq.~\eqref{eq:FBI} we obtain
\begin{equation}\label{eqs: etaBCs}
I_{\vec{\nu}}^{\Delta}(\mathbf{X},\eta) = \frac{(-1)^\nu\Gamma(\nu-\Delta)}{\prod_{i=1}^B \Gamma(\nu_b)}\eta^{\Delta-\nu} \left (1+\mathcal{O}(\eta^{-1})\right),
\end{equation}
where $\nu_b=\sum_{k=1}^{n_b}\nu_{(b,k)}$. With the boundary condition, the differential equations in Eq.~\eqref{eq:etaDE} can be solved following the AMFlow method \cite{Liu:2017jxz}. This allows us to obtain series expansions of $I_{\vec{\nu}}^{\Delta}(\mathbf{X},\eta)$ at $\eta=0$, $\eta=\infty$ and regular points on the negative imaginary axis. The expansion at $\eta=\infty$ is straightforward, which takes the form
\begin{equation}\label{eq
expansion}
I_{\vec{\nu}}^{\Delta}(\mathbf{X},\eta) = \eta^{\Delta-\nu}\sum_{i=0}^n c_i(\Delta,\mathbf{X},\vec{\nu}) \eta^{-i}.
\end{equation}
For expansions at regular points, the result is simply a Taylor series. However, the expansion at $\eta=0$, which may be a singular point,  generally takes the form
\begin{equation}
I_{\vec{\nu}}^{\Delta}(\mathbf{X},\eta) = \sum_{i}\eta^{\mu_i}\sum_j\log^j{\eta}\sum_{k=0}^n c_{ijk}(\Delta,\mathbf{X},\vec{\nu}) \eta^k.
\end{equation}

After expressing FBIs with auxiliary mass in terms of piecewise series expansions, the DCT integral Eq.~\eqref{eq:DCT} can be performed term by term straightforwardly. This can be done either for a given value of space-time dimension $D$, which gives a specific value for $\Delta$, or for an $\epsilon$-Laurent series in the case where $D=4-2\epsilon$. since $\Delta-\delta$ is fixed, the $\epsilon$-dependence is encoded in $\eta^{\delta-1}$ and $\Gamma(\delta)$, both of which can be expanded before  performing the integration. As a result, the computational cost of obtaining the expansion in $\epsilon$ is very low.

We have implemented the method in an open-source C++ package, which can compute one-loop FIs and FBIs to any desired order in $\epsilon$. Exceptional cases, such as vanishing Gram or modified Cayley determinants, are also handled.  The package is available at the following link:
\[
\href{https://gitlab.com/multiloop-pku/dct}{\text{https://gitlab.com/multiloop-pku/dct}}	
\]
Using template in C++, the program can work with any type of complex floating-point number representation. For precision goals beyond double precision, we utilize the boost::mpc data structure from the C++ Boost.Multiprecsion library \cite{BoostMultiprecision}, which relies on GMP \cite{GMP}, MPFR \cite{MPFR} and MPC \cite{MPC}.

\sect{Examples}
In the following, we demonstrate our approach by applying it to several examples. All calculations were performed using an Intel Core i7-8700K CPU. Please note that the reported numbers should be interpreted with caution, as computation time depends on the computer's workload. We choose a non-integer value for \(D_0 = \frac{7}{13}\). In practice, the value of \(D_0\) should not be too small (e.g., \(D_0 = -10 + \frac{7}{13}\)), as this would require more expansion terms to achieve a given accuracy. Similarly, \(D_0\) should not be too large (e.g., \(D_0 = 10 + \frac{7}{13}\)), as this could lead to precision cancellation during polynomial multiplication involving \(\epsilon\) in Eq.~(\ref{eq:DCT}).

\begin{figure}[htb]
	\begin{center}        \includegraphics[width=0.75\linewidth]{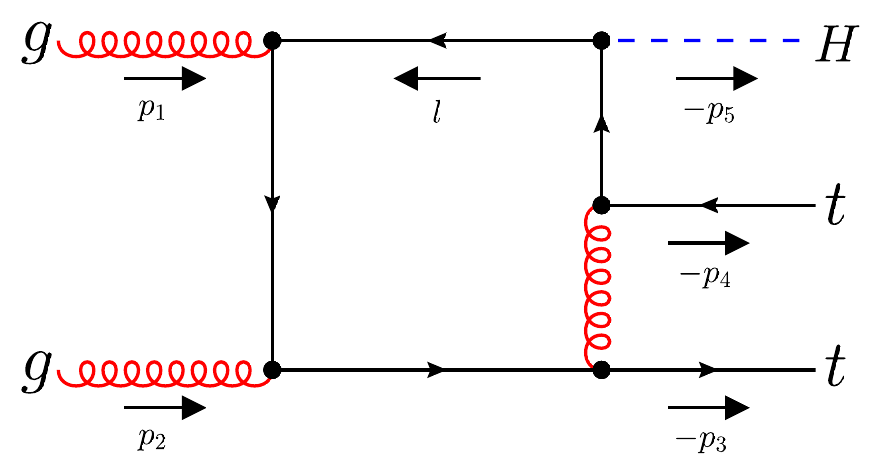}
	\end{center}	\caption{\label{fig:pentagon}A pentagon diagram in $gg\to t\bar{t}H$ process.}
\end{figure}

The first is an one-loop five-point example illustrated in Fig.~\ref{fig:pentagon}. This integral involves multiple mass scales which are relevant for the $gg\to t\bar{t}H$ process, corresponding to Topology C in Ref.~\cite{Buccioni:2023okz}.  The kinematics of this graph are characterized by Mandelstam variables $s_{ij}=(p_i+p_j)^2$ and on-shell conditions $p_1^2=p_2^2=0, p_3^2=p_4^2=m_t^2, p_5^2=m_H^2$, with all kinematic invariants expressible in terms of the set $\vec s=\{m_t^2,m_H^2,s_{12},s_{23},s_{34},s_{45},s_{51}\}$. We select a point $\vec s_0=\{1,\frac{1823191}{3567300},\frac{751}{23},-\frac{155}{11},\frac{85}{12},\frac{12271}{759},-\frac{59689}{3300}\}$ within the physical scattering region. 
During solving the $\eta$ differential equations \eqref{eq:etaDE}, we perform series expansions at 18 regular points and 2 singular points 0 and $\infty$, covering the entire negative imaginary axis. Employing DCT, we express 29 MIs in $D=4-2\epsilon$ as Laurent series 
 of $\epsilon$, up to $O(\epsilon^{10})$. We use build-in double precision and 
validate the correctness of our results by comparing it with the results obtained from the {\tt AMFlow} package \cite{Liu:2022chg}. In Tab.~\ref{table: precision}, we list the precision of top sector FI in the pentagon family.
\begin{table}[htb]
\centering
\begin{tabular}{| c c c c c c c c c c c c c|} 
\hline
$\epsilon$ order & -1 & 0 & 1 & 2 & 3 & 4 & 5 & 6 & 7 & 8 & 9 & 10    \\
\hline
precision & 15 & 15 & 14 & 14 & 14 & 14 & 15 & 14 & 14 & 14 & 15 & 14 \\
\hline
\end{tabular}
\caption{Precision of Laurent expansion coefficients of $I^{4-2\epsilon}_{({1,1,1,1,1})}$ defined in Fig.~\ref{fig:pentagon} with respect to $\epsilon$, where the precision is defined as $-\log_{10}(\varepsilon)$ and $\varepsilon$ is relative error. Double precision is used in the program.}
\label{table: precision}
\end{table}

\begin{figure}[htb]
	\begin{center}        \includegraphics[width=1.0\linewidth]{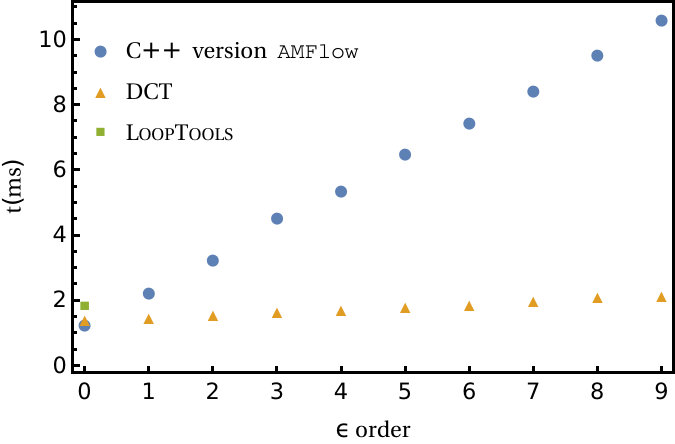}
	\end{center}	\caption{\label{fig:compare}Time spent for computing a pentagon family with increasing $\epsilon$ order by C++ version {\tt AMFlow}, DCT method and {\sc{LoopTools}}. The {\sc{LoopTools}} only computes the integrals to $O(\epsilon^0)$.}
\end{figure}

To demonstrate the efficiency of DCT, we also solve the integrals using {\sc{LoopTools}} and an in-house C++ version {\tt AMFlow}, where the integrals are solved directly in $D=4-2\epsilon$ dimensions by expanding the diferential equations and boundary conditions in $\epsilon$.  The comparisons are made using our C++ implementation in double precision. Note that the in-house C++ version {AMFlow} is much more efficient than the {\sc Mathematica} version {\tt AMFlow} package \cite{Liu:2022chg} for the computation of one-loop FIs. In Fig.~\ref{fig:compare}, we plot the wall time for evaluating a pentagon family to increasing orders of $\epsilon$ using all methods. Note that for evaluation up to $\epsilon^0$, all methods take less than 2 milliseconds, and the computation time increases linearly with $\epsilon$ for { AMFlow} method and  DCT. However, { AMFlow} exhibits a steeper increase. This is because solving DEs takes significantly longer time compared to performing the dimension-changing transformation. Note that the  {\sc{LoopTools}} only computes the integrals to $O(\epsilon^0)$.

\begin{figure}[htb]

	\begin{center}        \includegraphics[width=0.75\linewidth]{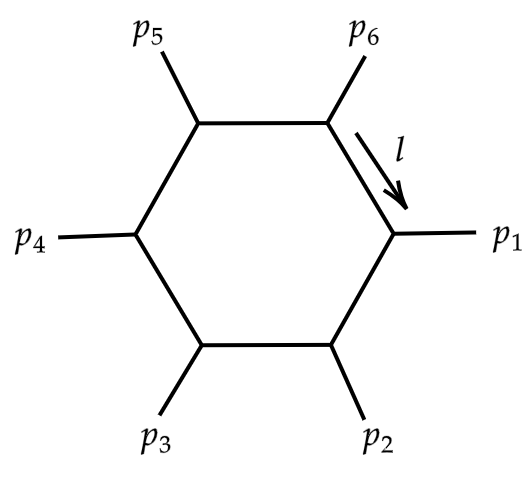}
	\end{center}	\caption{\label{fig:hexagon}A hexagon diagram. All propagators and external legs are massless.}
\end{figure}

We also demonstrate our approach with a massless one-loop six-point integral illustrated in Fig.~\ref{fig:hexagon}, which also been studied in Ref. \cite{Henn:2022ydo}. In this case where all propagators and external legs are massless, nine independent Mandelstam invariants can be chosen as $\vec{s}=\{s_{12}, s_{23}, s_{34}, s_{45}, s_{56}, s_{61}, s_{123}, s_{234}, s_{345}\}$. We select a point $\vec{s}_{0}=\{-1,-1,-1,-1,-1,-1,-\frac{1}{2},-\frac{5}{8},-\frac{17}{14}\}$, corresponding to a vanishing Gram determinant. We compute the integral at dimension $D=4-2\epsilon$. We use the boost::mpc type with 1000-digit precision and it takes 107 seconds to yield the results up to $\mathcal{O}(\epsilon^{10})$ for 51 MIs in this family. We validate the correctness of our results by comparing top sector FI with the results obtained from {\tt AMFlow} in Tab.~\ref{table: hex precision}.
\begin{table}[htb]
\centering
\begin{tabular}{| c c c c c c c c c c c c c c|} 
\hline
$\epsilon$ order & -2 & -1 & 0 & 1 & 2 & 3 & 4 & 5 & 6 & 7 & 8 & 9 & 10    \\
\hline
 precision-1000 & -3 &	-5 & -4 & -3 & -3 & -3 & -3 & -2 & -3 & -2 & -3 & -2 & -3\\
\hline
\end{tabular}
\caption{Precision of Laurent expansion coefficients of $I^{4-2\epsilon}_{({1,1,1,1,1,1})}$  defined in Fig.~\ref{fig:hexagon} with respect to $\epsilon$, where the precision is defined as $-\log_{10}(\varepsilon)$ and $\varepsilon$ is relative error. 1000-digit boost::mpc type is used in the program.}
\label{table: hex precision}
\end{table}

\begin{figure}[htb]

	\begin{center}        \includegraphics[width=0.75\linewidth]{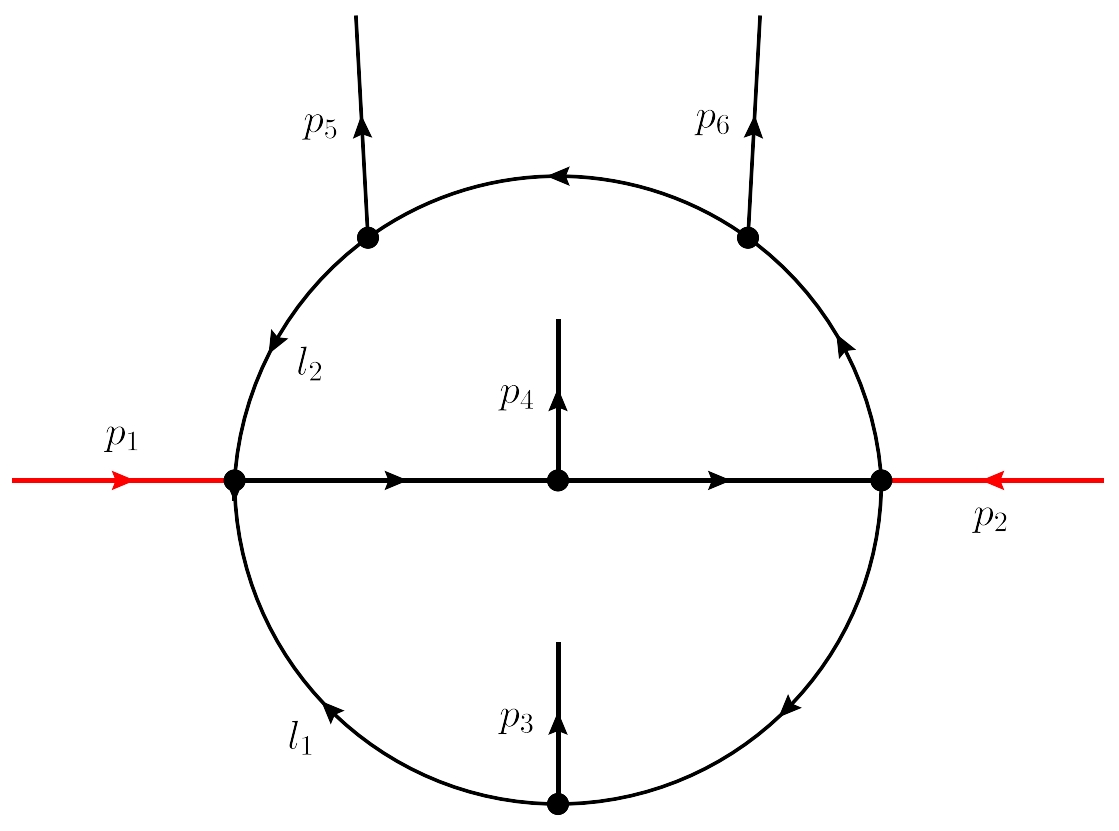}
	\end{center}	\caption{\label{fig:FBI}A two-loop diagram with three branches, red lines are massive and black lines are massless.}
\end{figure}

Finally, we demonstrate the effectiveness of the DCT method for computing FBIs through a two-loop six-point example, as shown in Fig.~\ref{fig:FBI}, which comprises 43 MIs. Under the kinematic configuration  \text{$p_{1,2}^2=1$}, \text{$p_{3,4,5,6}^2=0$} and $\{s_{12},s_{13},s_{14},s_{15},s_{23},s_{24},s_{25},s_{34},s_{35}\}=\{5,\frac{1}{3},\frac{1}{4},\frac{1}{5},\allowbreak \frac{1}{11},\frac{1}{7},\frac{1}{13}, \frac{1}{15},\frac{1}{17}\}$, we compute the entire family at $\mathbf{X}=\{\frac{1}{3},\frac{1}{2}\}$. By utilizing the double-precision DCT program, we obtained the Laurent expansion of the entire family up to $\mathcal{O}(\epsilon^5)$  within 2 milliseconds. As a comparison, for 30 values of $\epsilon$ ranging from $1\times 10^{-5}$ to $3\times 10^{-4}$, we established the differential equations of FBIs with respect to $\eta$. By numerically solving the differential equations using {\tt AMFlow} and fitting the coefficients of the $\epsilon$ expansion, we compared the calculation from DCT with this result, achieving a precision of 15 significant digits.

\sect{Summary and Outlook}
In this paper, we presented a novel method for computing one-loop Feynman integrals and fixed-branch integrals at any loop level to the desired order in \(\epsilon\). Our approach involves solving differential equations in a nonsingular space-time dimension using the auxiliary mass flow method, followed by applying a dimension-changing transformation to reach the target dimension, such as \(D = 4 - 2\epsilon\). This method is more efficient than the original auxiliary mass flow approach when computing to high orders in \(\epsilon\). The method has been implemented in an open-source C++ package, {\tt DCT}. We applied the package to several examples to demonstrate the validity and efficiency of our method.

We note that, while our current dimension-changing relations are achieved through integration over the auxiliary mass parameter \(\eta\), alternative dimension-changing relations also exist through integration over other parameters. For example, by performing the Laplace transform of \(\eta\)-space FIs in Eq.~(\ref{eqs: definition of I with eta}), we obtain a new quantity:
\[
W^{\Delta}_{\vec{\nu}}(\lambda) \equiv \int_0^\infty \mathrm{d}\eta\, e^{-\lambda \eta} I^{\Delta}_{\vec{\nu}}(\eta),
\]
which also results in a dimension-changing transformation:
\[
I^{\Delta}_{\vec{\nu}} = \frac{1}{\Gamma(\delta/2)\Gamma(1-\delta/2)} \int_0^\infty \mathrm{d}\lambda\, \lambda^{-\delta/2} W^{\Delta-\delta}_{\vec{\nu}}(\lambda).
\]
One can use similar techniques presented in this paper to compute FIs using this dimension-changing transformation, although \(\lambda = \infty\) is now an essential singularity that may introduce numerical difficulties.

\begin{acknowledgments}
	We would like to thank B. Feng for fruitful discussion. The work was supported by the National Natural Science Foundation of
	China (No. 12325503), the National Key Research and Development Program of China under
	Contracts No. 2020YFA0406400, the computing facilities at Chinese National Supercomputer Center in Tianjin and the High-performance Computing Platform of Peking University.
\end{acknowledgments}

\input{output.bbl}

\end{document}

%% file: output.bbl
\providecommand{\href}[2]{#2}\begingroup\raggedright\endgroup